\documentclass[twoside]{ima} 
\usepackage{newicktree}

\usepackage{amsmath}
\usepackage{amsfonts}
\usepackage{graphicx}
\usepackage{url}
\newtheorem{question}[theorem]{Problem}

\newcommand{\rr}{\mathbb R}
\newcommand{\zz}{\mathbb Z}
\newcommand{\A}{\texttt{A}}
\newcommand{\C}{\texttt{C}}
\newcommand{\G}{\texttt{G}}
\newcommand{\T}{\texttt{T}}

\newcommand{\bfT}{\mathbf{f}_T}
\newcommand{\hpt}{\hat{p}(\theta)}

\newcommand{\onet}{\frac {1} {3}}

\newcommand{\tr}{\mathrm{tr}}
\newcommand{\bof}{\mathbf{f}}
\begin{document}
\markboth{NICHOLAS ERIKSSON}%
{USING INVARIANTS FOR PHYLOGENETIC TREE CONSTRUCTION} 

\title{USING INVARIANTS FOR PHYLOGENETIC TREE CONSTRUCTION} 
\author{NICHOLAS ERIKSSON\thanks{Department of Statistics,
	University of Chicago, Chicago, IL 60637,
	\texttt{eriksson@galton.uchicago.edu}, 
	partially supported by the NSF (DMS-06-03448)}}

\maketitle   
\begin{abstract}
	Phylogenetic invariants are certain polynomials in the joint probability
	distribution of a Markov model on a phylogenetic tree.  Such polynomials
	are of theoretical interest in the field of algebraic statistics and they
	are also of practical interest---they can be used to construct
	phylogenetic trees.  This paper is a self-contained introduction to the
	algebraic, statistical, and computational challenges involved in the
	practical use of phylogenetic invariants.  We survey the relevant
	literature and provide some partial answers and many open problems.
\end{abstract}


{\AMSMOS 92B10, 92D15, 13P10, 05C05 \endAMSMOS}

\section{Introduction} \label{sec:intro}

The emerging field of algebraic statistics (cf.\ \cite{Pachter2005}) has at its
core the belief that many statistical problems are inherently algebraic.
Statistical problems are often analyzed by specifying a \emph{model}---a
family of possible probability distributions to explain the data.  In
particular, many statistical models are defined parametrically by polynomials
and thus involve algebraic varieties.  From this point of view, one would hope that
the ideal of polynomials that vanish on a statistical model would give
statistical information about the model.  This is not a new idea in statistics,
indeed, tests based on polynomials that vanish on a model include the
\emph{odds-ratio}, which is based on the determinant of a two by two matrix. 
The polynomials that vanish on the statistical model have come to be known as
the \emph{(algebraic) invariants} of the model.

The field of phylogenetics provides important 
statistical and biological models with interesting
combinatorial structure.
The central problem in
phylogenetics is to determine the evolutionary relationships
among a set of \emph{taxa} (short for taxonomic units, which could be
species, for example).  To a first approximation, these relationships can be
represented using rooted binary trees, where  the leaves correspond to the
observed taxa and the interior nodes to ancestors.  For example,
Figure~\ref{fig:1} shows the relationships between a portion of a gene in seven
mammalian species.

Phylogenetic invariants are polynomials in the joint probability distribution
describing sequence data that vanish on distributions arising from a
particular tree and model of sequence evolution.  The first of
the invariants for phylogenetic tree models were discovered by Lake
and Cavender-Felsenstein \cite{Lake1987,Cavender1987}.  
This set off a flurry of
work: in mathematics, generalizing these invariants (cf.\
\cite{Hendy1989,Evans1993, Szekely1993}) and in phylogenetics, using these
invariants to construct trees 
(cf.\ \cite{ Holmquist1988, Navidi1991, Navidi1993, Sankoff1998, Sankoff2000}).
However, the linear invariants didn't fare well in simulations
\cite{Huelsenbeck1995} and the idea fell into disuse.  

However, the study of phylogenetic invariants was revived in the field
of algebraic statistics; the subsequent theoretical (cf.\
\cite{Allman2003,Sturmfels2005,Casanellas2007,Allman2007c}) and practical
(cf.\ \cite{Casanellas2005, Casanellas2006, Eriksson2005, Eriksson2007,
Kim2006}) developments have 
given cause for optimism in using invariants to construct phylogenetic trees.
There are benefits to these  algebraic tools; however, obstacles in algebraic geometry,
statistics, and computer science must be overcome if they are to live up to
their potential.  In this paper, we formulate and analyze some of the
fundamental advantages and difficulties in using algebraic statistics to
construct phylogenetic trees, describing the current research and formulating many open problems.

In geometric terms, the problem of phylogenetic tree construction can be stated as 
follows.  We observe DNA sequences from $n$ different taxa and wish to
determine which binary tree with $n$
leaves best describes the relationships between these sequences for a fixed model of evolution.  
Each of these
trees corresponds to a different 
algebraic variety in $\rr^{4^n}$.  
The DNA sequences correspond to a certain point in $\rr^{4^n}$ as well.
Picking the best tree means picking the variety that is closest to the data
point in some sense.   
Since the data will not typically lie on the variety
of any tree,  we have to decide what is meant by ``close''.
	
Denote the variety (resp.\ ideal) associated
to a tree $T$ by $V(T)$ (resp.\ $I(T)$).
Our main goal, then, is to understand how the polynomials in $I(T)$ can be used
to select the best tree given the data.  In order to answer this
question, there are five  fundamental obstacles. 
\begin{enumerate} 
	\item Formulate an appropriate model of evolution and determine the
		invariants for that model, if possible in a form that can be evaluated
		quickly. 
	\item Choose a finite set of polynomials in $I(T)$ with good discriminating
		power between different trees.  
	\item Given a set of invariants for each tree, define a single score that
		can be used to compare different trees.
	\item Since the varieties are in $\rr^{4^n}$, each polynomial is in
		exponentially many unknowns.  Thus even evaluating a single invariant
		could become difficult as $n$ increases.   This is in addition to the
		problem that the number of trees and the codimension of $V(T)$ increase
		exponentially.  Phylogenetic algorithms are often used for hundreds of
		species.  Can invariants become practical for large problems?
	\item Statistical models are not complex algebraic varieties; they make
		sense only in the probability simplex and thus are real, semi-algebraic
		sets.  This problem is more than theoretical---it is quite noticeable
		in simulated data (see Figures~\ref{fig:dist} and \ref{fig:distR}).
		Can semi-algebraic information be used to augment the invariants?  
\end{enumerate}

In the remainder of the paper, we will analyze these problems in detail,
showing why they are significant and explaining  some methods for dealing with
them.  The first problem (determining phylogenetic invariants) has been the
focus of substantial research, thus we deal here with only the last four
problems.  We begin by introducing phylogenetics and constructing and using
some phylogenetic invariants, then consider the four problems in order.

While in this paper we concentrate solely on the problem of constructing
phylogenetic trees using invariants, we should note that phylogenetic
invariants are interesting for many other reasons.  On the theoretical side of
phylogenetics, they have been used to answer questions about identifiability
(e.g., \cite{Allman2006,Matsen2007}).  The study of the algebraic geometry
arising from invariants has led to many interesting problems in mathematics
\cite{Eriksson2005,Buczynska2007,Cox2007}.

\section{Background} \label{sec:back}

We give here a short, self-contained introduction to phylogenetics and
phylogenetic invariants.  For a more thorough survey of algebraic methods in
phylogenetics, see \cite{Allman2007}.  Also see \cite{Felsenstein2003,Semple2003}
for more of the practical and combinatorial aspects of phylogenetics.

\begin{definition} \rm
	Let $X$ be a set of taxa.
	A \emph{phylogenetic tree} $T$  on $X$ is a unrooted binary tree with $|X|$
	leaves where each leaf is labelled with an element of $X$ and each edge $e$ of
	$T$ has a weight, written $t_e$ and called the \emph{branch length}. 
\end{definition}

While we include branch lengths in our definition of phylogenetic trees, our
discussions about constructing trees are about only choosing the correct
topology (meaning the topology of the labelled tree), not the branch lengths.  
While estimating branch lengths is
relatively easy using maximum likelihood methods after a tree topology is fixed
(e.g., with \cite{Yang1997}), it is an interesting question whether algebraic ideas
can be used to estimate branch lengths (see \cite{Steel2000, Allman2008} for 
algebraic techniques for estimating parameters in invariable-site phylogenetic models).

\begin{figure}
	\begin{center}
		\begin{newicktree}
			\righttree
			\setunitlength{10cm}
			\drawtree{(Armadillo:0.17,(Dog:0.15,(Human:0.14,
(Rabbit:0.19,(Mouse:0.08,Rat:0.12):0.31):0.10):0.07):0.03,
Elephant:0.21);}
		\end{newicktree}
	\end{center} 
	\caption{Phylogenetic tree for seven mammalian species derived
	from an alignment of a portion of the \emph{HOXA} region (ENCODE region
	ENm010, see \cite{Consortium2004} and \texttt{genome.ucsc.edu/encode}).
	This tree was built using the \texttt{dnaml} maximum likelihood package from \texttt{PHYLIP} \cite{Felsenstein2005}
	on an alignment partially shown in Figure~\ref{fig:msa}.}
	\label{fig:1}
\end{figure}

\begin{figure}
	\centering
	\includegraphics[width=4in]{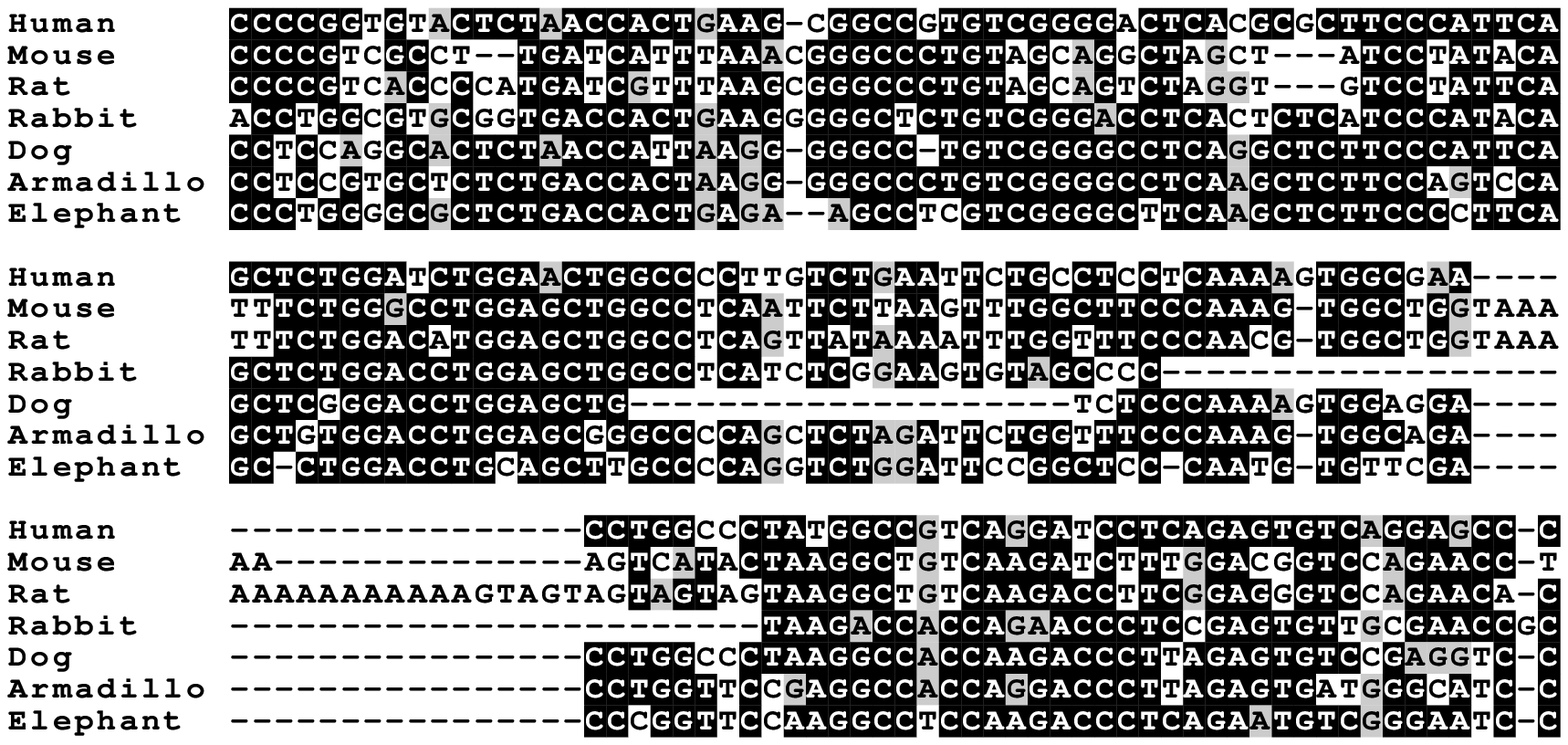} 
	\caption{Multiple sequence alignment of length 180 from the HOXA region of
	seven mammalian genomes.  Dashes indicate gaps; bases are colored according
	to their similarity across the species.}
	\label{fig:msa}
\end{figure}

Phylogenetics depends on having identified \emph{homologous characters} between
the set of taxa.  For example, historically, these characters might be physical
characteristics of the organisms (for example, binary characters might include
the following: are they unicellular or multicellular, cold-blooded or
hot-blooded, egg-laying or placental mammals).  In the era of genomics, the characters
are typically single nucleotides or amino acids that have been inferred to be
homologous (e.g., the first amino acid in a certain gene that is shared in a
slightly different form among many organisms).  For example, see
Figure~\ref{fig:msa}, which shows a multiple sequence alignment.  We will
throughout make the typical assumption that characters evolve independently, so
that each column in Figure~\ref{fig:msa} is an independent, identically distributed (i.i.d.) 
sample from the model
of evolution.  While both DNA and amino acid data are common, we will work only
with DNA and thus use the alphabet $\Sigma = \{ \A, \C , \G, \T \}$.

We assume that evolution happens via a continuous-time Markov process on a
phylogenetic tree (see \cite{Norris1997} for general details about Markov
chains).  That is, along each edge $e$ there is a length $t_e$ 
 and a rate matrix $Q_e$ giving the instantaneous rates for
evolution along edge $e$.  Then $M_e = e^{Q_e t_e}$  is the transition matrix
giving the probabilities of substitutions along the edge.  In order to work with
unrooted trees, we will assume that the Markov process is reversible, that is,
$\pi_i M_e(i,j) = \pi_j M_e(j,i)$, where $\pi$ is the stationary distribution
of $M_e$.  In order for $e^{Q_e t_e}$ to be stochastic, we must have $Q(i,i) \leq 
0$, $Q(i,j) \geq 0$ for $i \neq j$, and $\sum_{j} Q(i,j) = 0$ for all $i$.
Notice that since $\det(e^{Q}) = e^{\tr(Q)}$, we can recover the  branch length
from the transition matrix $M_e$ as
\begin{equation}
	\label{eq:logdet}
	t_e = \frac 1 {\tr Q_e} \log \det(M_e).
\end{equation}

\begin{example} \rm
	\label{ex:jc}
	Let 
	$Q_e = \begin{pmatrix}
		-1 & \onet & \onet & \onet\\
		\onet & -1 & \onet & \onet\\
		\onet & \onet & -1 & \onet\\
		\onet & \onet & \onet & -1 
	\end{pmatrix}
	$
	be the rate matrix for edge $e$, 
	where the rows and columns are labeled by $\Sigma = \{ \A, \C, \G, \T \}$.
	Then 
	\[
	M_e = e^{Q_et_e} = \frac 1 4 
	\begin{pmatrix}
		1 + 3 e^{-\frac{4}{3}t_e} & 1 - e^{-\frac{4}{3}t_e}& 
		1 - e^{-\frac{4}{3}t_e}& 1 - e^{-\frac{4}{3}t_e}\\ 
		1 - e^{-\frac{4}{3}t_e}& 1 + 3 e^{-\frac{4}{3}t_e} & 
		1 - e^{-\frac{4}{3}t_e}& 1 - e^{-\frac{4}{3}t_e}\\ 
		1 - e^{-\frac{4}{3}t_e}& 1 - e^{-\frac{4}{3}t_e}& 
		1 + 3 e^{-\frac{4}{3}t_e} & 1 - e^{-\frac{4}{3}t_e}\\ 
		1 - e^{-\frac{4}{3}t_e}& 1 - e^{-\frac{4}{3}t_e}& 
		1 - e^{-\frac{4}{3}t_e}& 1 + 3 e^{-\frac{4}{3}t_e} 
	\end{pmatrix}.
	\]
	This form of rate matrix is known as the Jukes-Cantor model \cite{Jukes1969}.
	For example,  the probability of changing from an \A\ to a \C\ along edge $e$
	is given by $M_e(1,2) = \frac{1 - e^{-\frac{4}{3}t_e}}{4}$.
\end{example}

Commonly used models that are more realistic than the Jukes-Cantor model include the
Kimura 3-parameter model \cite{Kimura1981} where the rate matrices are of the form
\[
\begin{pmatrix}
        \cdot & \gamma & \alpha & \beta\\
        \gamma & \cdot & \beta & \alpha\\
        \alpha & \beta & \cdot & \gamma\\
        \beta & \alpha & \gamma & \cdot
\end{pmatrix},
\]
where $\cdot = -\gamma - \alpha - \beta$.
See \cite[Figure~4.7]{Pachter2005} for a description of many other possible models.

In order to obtain the joint distribution of characters at the
leaves of the trees, we have to choose a root of the tree (arbitrarily, since the
processes are time reversible), and run the Markov process down the edges of the
tree, starting from a distribution of the characters at the root.
The result is  a joint probability distribution $p = (p_{\A \dots \A},
\dots, p_{\T \dots \T})$, and the important point is that the coordinates of
$p$ can be written as polynomials in the transition probabilities.  That is,
the model is specified parametrically by polynomials in the entries of $M_e$.
We will forget about the specific form of the entries of $M_e = e^{Q_e t_e}$
and instead treat each entry of $M_e$ as an unknown.  Thus for the Jukes-Cantor
model, we have two unknowns per edge: $\alpha_e = \frac {1 + 3 e^{- \frac {4}
{3} t_e}} {4}$ and $\beta_e = \frac {1 - e^{- \frac {4} {3} t_e}} {4}$.  
This makes the algebraic model more general than the statistical model (as it
allows probabilities in the transition matrices to be negative or even
complex).  Although this allows algebraic tools to be used,  we will see in
Section~\ref{sec:pos} that it can be a disadvantage.
Generally speaking, there are two types of phylogenetic models that have been
thoroughly studied from the algebraic viewpoint: ``group based'' models such
as the Jukes-Cantor and Kimura models, and variants of the general Markov model, in which no
constraints are placed on the transition matrices.

\begin{figure}
	\centering
	\includegraphics[width=\textwidth]{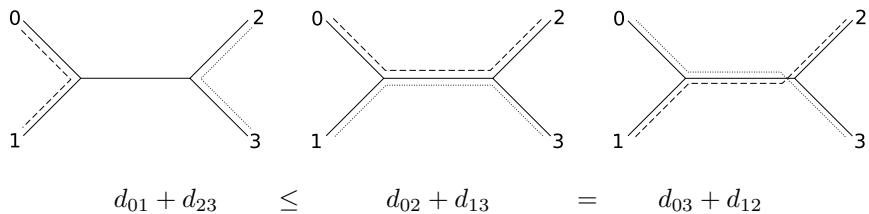}
	\[ 
	d_{01} + d_{23} \qquad \leq \quad \qquad d_{02} + d_{13}\quad  \qquad = \qquad d_{03} + d_{12}
	\]
	\caption{The four-point condition.}  
	\label{fig:4pt}
\end{figure}

In this paper, we define the \emph{phylogenetic invariants} for a model of
evolution and a tree to be the polynomials in the joint probabilities that 
vanish if the probabilities come from the model on the tree.  For example, for
a quartet tree (an unrooted binary tree with four leaves, see Figure~\ref{fig:4pt}) 
under the Jukes-Cantor model, $p_{\A\A\A\A} - p_{\C\C\C\C} = 0$,
due to the symmetry built into the model.  However, this polynomial doesn't
differentiate between trees---it lies in the intersection of the ideals of the three
quartet trees.  Beware that there are two commonly used definitions of
phylogenetic invariants.  Originally, they were defined as polynomials that
vanish on probability distributions arising from exactly one tree, so the above
polynomial would be excluded.  However, it is more algebraically convenient to
take as invariants the full set of polynomials that vanish, as this forms an
ideal.  We spend the rest of this section deriving a particularly important
polynomial.

A class of phylogenetic methods bypass working with the joint probability
distribution and instead only estimate the distances between each pair of taxa.
The goal then is to find a tree with branch lengths such that the distance
along edges of the tree between pairs of leaves approximates the estimated
pairwise distances.  To use these distance methods, we first need a couple of
definitions.
We will concentrate in this paper on \emph{quartet} trees, i.e., 
trees with four leaves.  There are 3 different (unrooted, binary)
trees on four leaves, we will write them $(01 : 23)$, $(02 : 13)$, and $(03 :
12)$, corresponding to which pairs of leaves are joined together.

\begin{definition}
	A \emph{dissimilarity map} $d \in \rr^{\binom{n}{2}}$ satisfies $d(i,j) =
	d(j,i) \geq 0$ and $d(i,i) = 0$.  We say that $d$ is a \emph{tree metric}
	if there exists a phylogenetic tree $T$ with non-negative branch lengths
	$t_e$ such that for every pair $i,j$ of taxa, $d(i,j)$ is the sum of the
	branch lengths $t_e$ on the edges of $T$ connecting $i$ and $j$.
\end{definition}

\begin{proposition}[Four-point condition \cite{Buneman1974}]  \label{prop:4pt}
	A dissimilarity map $d$ is a tree metric if and only if for every $i, j, k,$ and 
	$l$, the maximum of the three numbers 
	\[ d_{ij} + d_{kl}, \quad  d_{ik} + d_{jl}, \text{ and} \quad d_{il} + d_{jk} \]
	is attained at least twice.
\end{proposition}

\begin{example} \label{ex:4pt} \rm
Let us restrict our attention to a tree with four leaves, $(ij : kl)$.  
In this case, the four-point condition becomes (see Figure~\ref{fig:4pt})
\begin{equation} \label{eq:4pt}
d_{ij} + d_{kl} \leq  d_{ik} + d_{jl} = d_{il} + d_{jk}.
\end{equation}
The equality in the four-point condition can be translated into a quadratic
polynomial in the probabilities, however, we first have to understand how to
transform the joint probabilities into distances.  Distances can be estimated
from data in a variety of ways (there are different formulas for the maximum likelihood 
estimates of the distances under different models of evolution, see
\cite[Chapter 13]{Felsenstein2003}).   The formula for the general Markov model
is the logdet distance, which mimics what we saw
above (\ref{eq:logdet}), in that a transition matrix is estimated and the
distance is taken to be the log of the determinant of this  matrix.

Here we will use a simpler formula for the distance, under the Jukes-Cantor model (Example~\ref{ex:jc}).
The maximum likelihood estimate of the distance between two sequences
under the Jukes-Cantor model is given by 
$d_{ij} = - \frac 1 4 \log \left( 1 - \frac {4 m_{ij}} {3}\right)$
where $m_{ij}$ is the fraction of mismatches between the two sequences, e.g.,
\[
m_{12} = \sum_{i,j,k,l \in \{\A,\C,\G,\T\}, i \neq j} p_{ijkl}
\]

After plugging this distance into the four point condition, cancelling, and
exponentiating, the equality in (\ref{eq:4pt}) becomes
\begin{equation} \label{eq:h10}
\left (1 - \frac {4} {3} m_{ik}\right) \left (1 - \frac {4} {3} m_{jl}\right) 
-
\left (1 - \frac {4} {3} m_{il}\right) \left (1 - \frac {4} {3} m_{jk}\right) = 0.
\end{equation}
We will call this polynomial the four-point invariant.
This construction is originally due to  Cavender and Felsenstein
\cite{Cavender1987}.
\end{example}

Example~\ref{ex:4pt} shows one of the first constructions on a phylogenetic
invariant, in the same year as the discovery by Lake of linear invariants
\cite{Lake1987}.  There is a linear change of coordinates on the probability
distribution $p$  under which $I(T)$ has a generating set of binomials.  In particular, in these
coordinates, a simple calculation shows that (\ref{eq:h10}) becomes a binomial.  Known as the
Hadamard or Fourier transform \cite{Hendy1989,Szekely1993,Evans1993,Sturmfels2005},
this change of coordinates transforms the ideals of invariants for several models of
evolution into toric ideals \cite{Sturmfels1996}. It should be emphasized, however, that
this transform is only known to exist for group-based models.

The four-point invariant is a polynomial in the joint probabilities that vanishes
on distributions arising from a certain quartet tree.  Define the ideal $I_{\mathcal M}(T)$ 
of invariants for a model $\mathcal M$ of evolution on a tree $T$  to be the
set of all polynomials that are identically zero on all probability
distributions arising from the model $\mathcal M$ on $T$.
We will write only $I(T)$ when $\mathcal M$ is clear from context.

\section{How to use invariants} \label{sec:howto}
The basic idea of using phylogenetic invariants is as follows.  A multiple
sequence alignment DNA alignment of $n$ species gives rise to an empirical 
probability distribution $\hat{p} \in \rr^{4^n}$.  This occurs simply by
counting columns of each possible type in the alignment, throwing out all
columns that contain a gap (a ``\texttt{-}'' symbol).  For example,
Figure~\ref{fig:msa} has exactly one column that reads ``\C\C\C\A\C\C\C'' (the
first) out of 107 gap-free columns total, so $\hat{p}_{\C\C\C\A\C\C\C} = 1/107$.

Then if $f$ is an invariant for tree $T$ under a certain model of evolution, 
we expect $f(\hat{p}) \approx 0$ if (and generically only if) the alignment
comes from the model on $T$.  More precisely,
where $\hat{p}_N$ is the empirical distribution after seeing $N$ observations
from the model on $T$, then 
\(
\lim_{N \to \infty} E(f(\hat{p}_N)) = 0.
\)

We thus have a rough outline of how to use phylogenetic invariants to construct
trees:
\begin{enumerate}
	\item Choose a model $\mathcal M$ of evolution.
	\item Choose a set of invariants  $\bfT$ for model $\mathcal M$ for each tree $T$ with $n$ leaves.
	\item Evaluate each set of invariants at $\hat{p}$.  
	\item Pick the tree $T$ such that $\bfT(\hat{p})$ is smallest (in some
		sense).
\end{enumerate}

However, all of these steps contain difficulties:  there are infinitely many
polynomials to pick in exponentially many unknowns and exponentially many trees
to compare.  
We will discuss step 2 in Section~\ref{sec:choose}, step 3 in
Section~\ref{sec:compute}, and step 4 in Section~\ref{sec:compare}.
Selecting a  model of evolution is difficult as  well.  There is, as always, a
trade-off between biological realism (which could lead to hundreds of parameters per edge)
and statistical usefulness of the model. 

Since the rest of this paper will discuss difficulties with using invariants,
we should stop and emphasize two especially promising features of invariants:

\paragraph{1. Invariants allow for arbitrary rate matrices.} One major
challenge of phylogenetics is that evolution does not always happen at one rate.
But common methods for constructing trees generally assume a single rate
matrix $Q$ for all edges, leading to difficulties on data with heterogeneous
rates.  
While methods have been developed to solve this problem (cf.\ \cite{Yang1995,
Galtier1998, Gascuel2007}), it is a major focus of research.

In contrast, phylogenetic invariants allow for differing rate matrices within
the chosen model on every edge (and in fact, even changing rate matrices along
a single edge).  The invariants for the Kimura 3-parameter model
\cite{Kimura1981} have been shown to outperform neighbor-joining and maximum
likelihood on quartet trees for heterogeneous simulated data
\cite{Casanellas2006}.  
To be fair, we should note that the invariants in this analysis were based on
the correct model (i.e., the Kimura 3-parameter with heterogeneous rates, which
the data was simulated from) while the maximum likelihood analysis used an
incorrect model (with homogeneous rates) due to limitations in standard maximum
likelihood packages.

\paragraph{2. Invariants perhaps can test individual features of trees.}
Researchers are frequently interested in the validity of a single edge in
the tree. For example, we might wonder if human or dog is a closer relative
to the rabbit.  This amounts to wondering about how much confidence there
is in the edge between the human-rabbit-mouse-rat subtree and the dog
subtree in Figure~\ref{fig:1}.  There are methods, most notably the bootstrap
\cite{Felsenstein1985} and Bayesian methods (cf.\ \cite{Huelsenbeck2001}),
which provide answers to this question, but there are concerns about their
interpretation \cite{Hillis1993,Efron1996,Newton1996,Alfaro2003}.

As for phylogenetic invariants, the generators of the ideal $I(T)$ are,
in many cases, built from polynomials constructed from local features
of the tree.  Thus invariants seem to be well suited  to test
individual features of a tree.  For example, suppose we have $n$ taxa.
Consider a partition $\{A, B\}$ of the taxa into two sets.  Construct
the $|\Sigma|^{|A|} \times |\Sigma|^{|B|}$ matrix
$\operatorname{Flat}_{A,B}(p)$ where the rows are indexed by
assignments of $\Sigma$ to the taxa in $A$ and the columns by
assignments of $\Sigma$ to the taxa in $B$.  The entry of the matrix
in a given row and column is the joint probability of seeing the
corresponding assignments of $\Sigma$ to $A$ and $B$.  The following theorem is
\cite[Theorem~4]{Allman2007b} and deals with the general Markov model, where
there are no conditions on the form of the rate matrices.  
\begin{theorem}[Allman-Rhodes] \label{thm:flat} Let $\Sigma = \{0, 1\}$ and
	let $T$ be a binary tree under the general Markov model.  Then the $3
	\times 3$ minors of $\operatorname{Flat}_{A,B}(p)$ generate $I(T)$ for the
	general Markov model, where  we let $A, B$ range over all partitions of
	$[n]$ that are induced by removing an edge of $T$.  
\end{theorem}

While the polynomials in Theorem~\ref{thm:flat} do not generate the ideal
for the DNA alphabet, versions of these polynomials do vanish for any Markov
model on a tree.
A similar result also holds for the Jukes-Cantor model in Fourier
coordinates;  the following is part of \cite[Thm~2]{Sturmfels2005}.
\begin{theorem}[Sturmfels-Sullivant] \label{thm:jc} 
	The ideal for the Jukes-Cantor DNA model is generated by polynomials of
	degree 1, 2, and 3 where the quadratic (resp.\ cubic) invariants are
	constructed in an explicit combinatorial manner from the edges (resp.\
	vertices) of the tree.
\end{theorem} 

Although there are many challenges to overcome, the fact that phylogenetic
invariants are associated to specific features of a tree provides hope that
they can lead to a new class of statistical tests for individual features on
phylogenetic trees.

\section{Choosing powerful invariants} \label{sec:choose}
There are, of course, infinitely many polynomials in each ideal $I(T)$, and it
is not clear mathematically or statistically which should be used in the set
$\bfT$ of invariants that we test.  For example, we might hope to use a generating set, or a Gr\"obner basis, or a set
that locally defines the variety, or a set that cuts out the variety over
$\rr$.  We have no actual answers to this dilemma, but we provide a few
illustrative examples and suggest possible criteria for an invariant to be
powerful.  We will deal with the Jukes-Cantor model on a tree with four leaves; 
the 33 generators for this ideal can be found on the 
``small trees'' website \url{www.shsu.edu/~ldg005/small-trees/} \cite{Casanellas2005}.

We believe that symmetry is an important factor in choosing powerful invariants.
The trees with four leaves have a very large symmetry group: each tree can be
written in the plane in eight different ways (for example, one tree can
be written as (01 : 23),  (10 : 23), \dots, (32 : 10)), and each of these
induces a different order on the probability coordinates $p_{ijkl}$.  
This symmetry group ($\zz_2 \times \zz_2 \times \zz_2$) acts on the ideal $I(T)$ as well.
In order that the results do not change under different orderings of the input, 
we should choose a set $\bfT$ of invariants that is closed (up to sign) under
this action.  After applying this action to the 33 generators, we get a set of 49 invariants.
This symmetry will also play an important role in our metric
learning algorithms in Section~\ref{sec:compare}.
See also \cite{Sumner2007} for a different perspective on symmetry in phylogenetics.

\begin{figure}
	\centering
	\begin{tabular}{ccc}
	\includegraphics[angle=270,width=.28\textwidth]{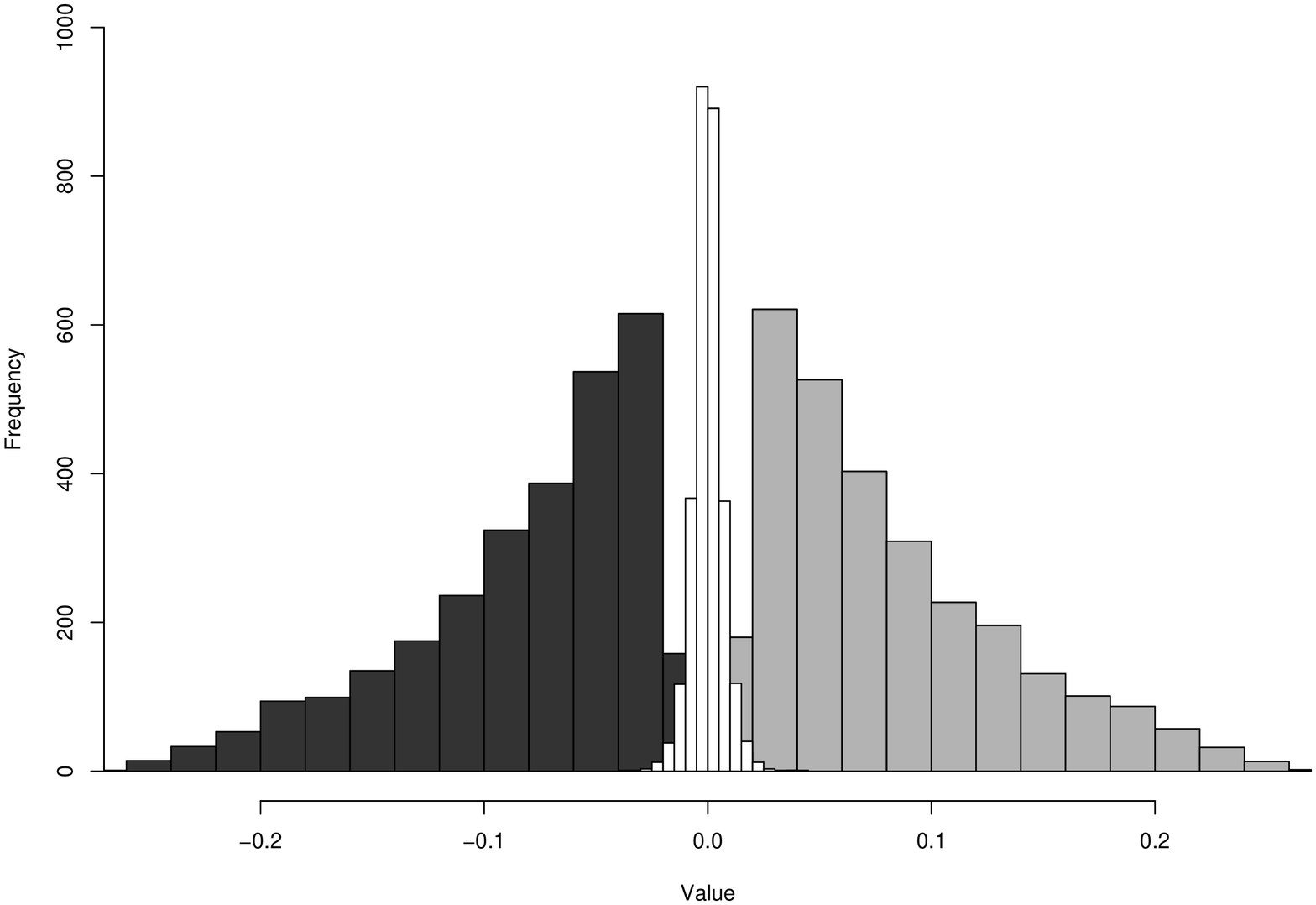} &
	\includegraphics[angle=270,width=.28\textwidth]{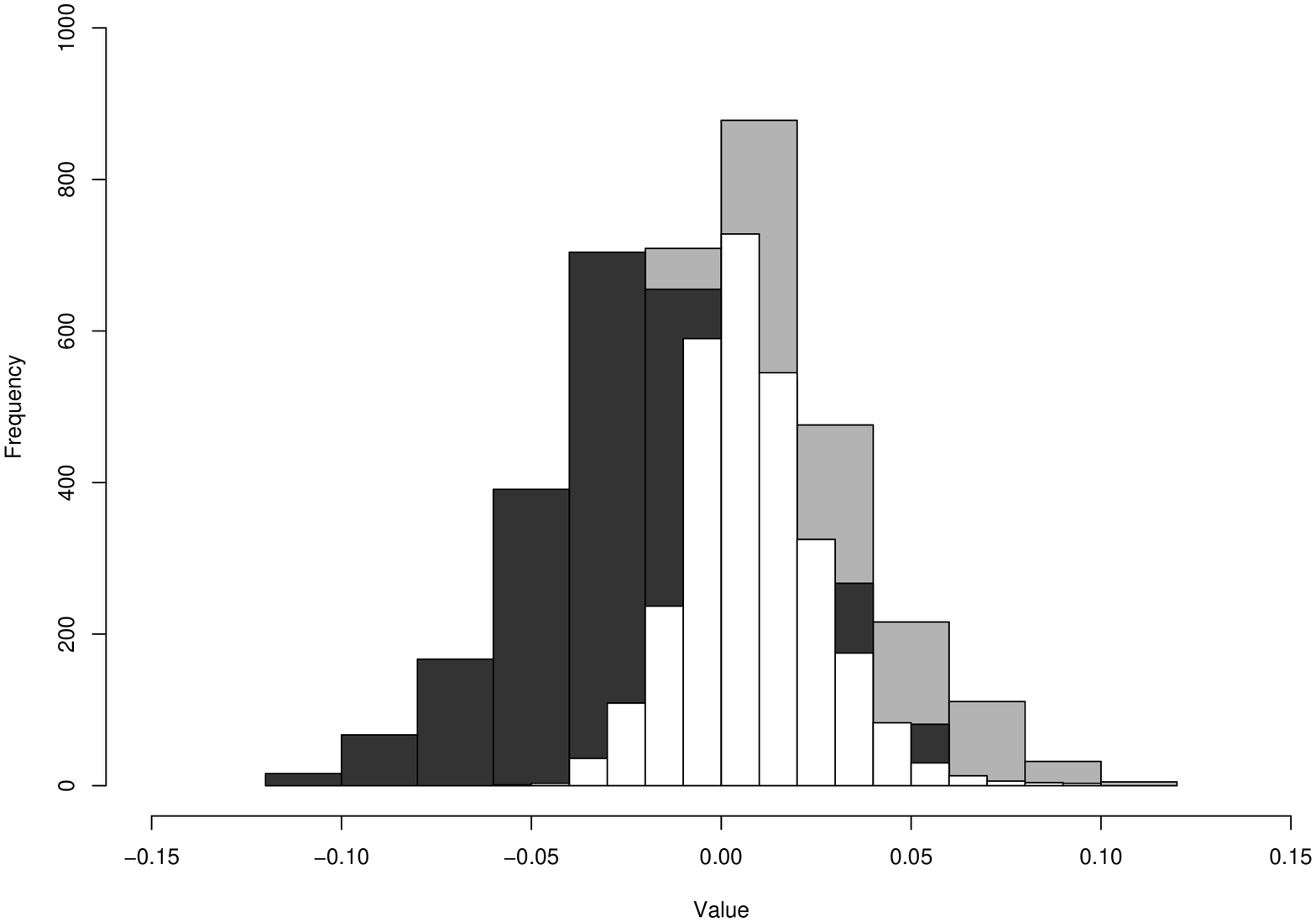} &
	\includegraphics[angle=270,width=.28\textwidth]{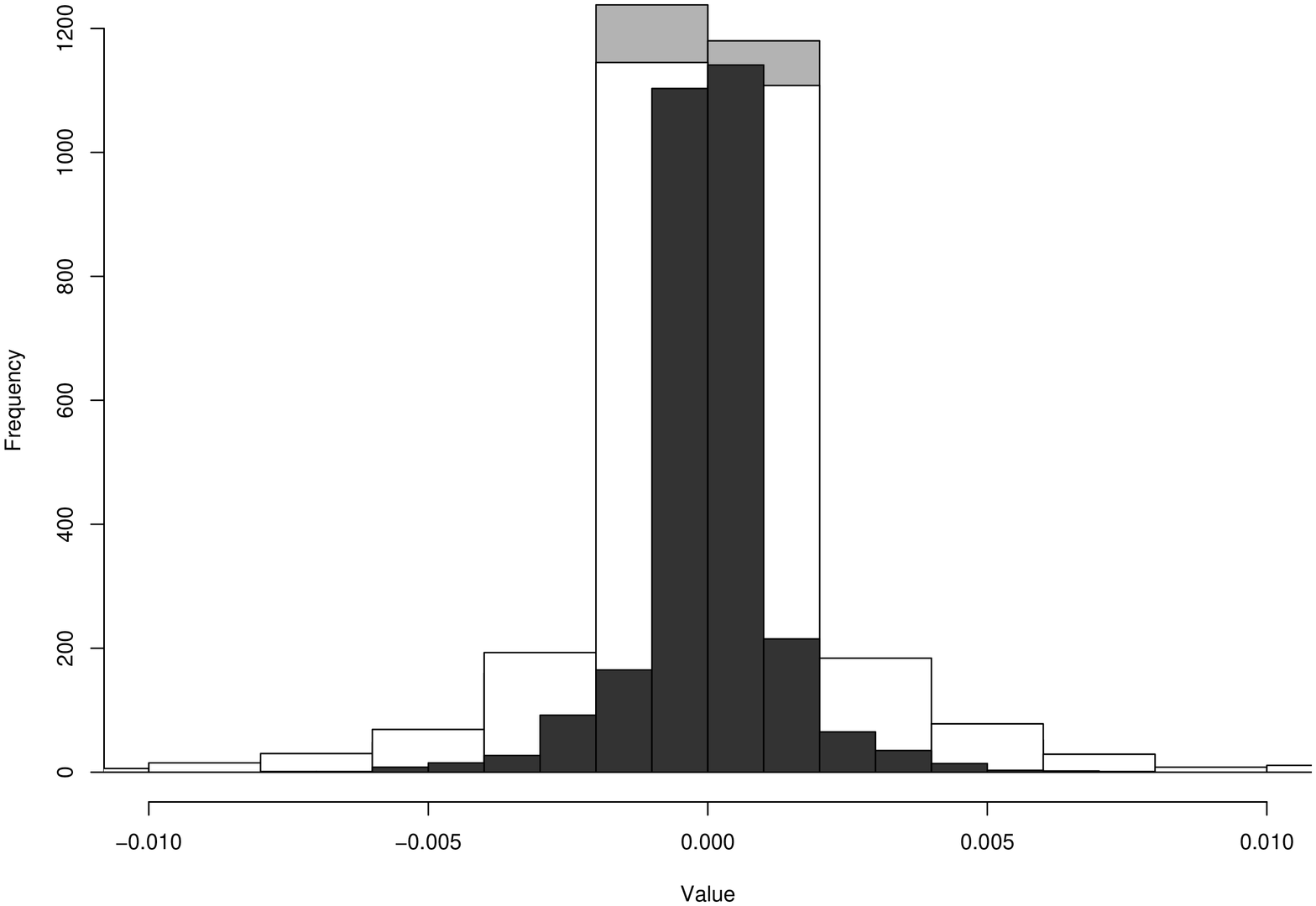}\\
	Four-point & Lake's & Biased
\end{tabular}
	\caption{
	Distributions of three invariants (the four-point invariant, Lake's linear
	invariant,  and a biased invariant) on simulated data.  The white 
	histogram corresponds to the correct tree, the black and gray are the other
	two trees.  The invariants have quite different variances and performance.}
	\label{fig:hist} 
\end{figure}

We begin by showing how different polynomials have drastically different behavior.
Figure~\ref{fig:hist} shows the distribution of three of the invariants on data
from simulations of alignments of length 1000 from the Jukes-Cantor model on $(01 :
23)$ for branch lengths ranging from $0.01$ to $0.75$ (similar to 
\cite{Huelsenbeck1995,Casanellas2006,Eriksson2007}).
The histograms show the distributions for the simulated tree in white 
and the distributions for the other trees in gray and black.
The four-point invariant (left) distinguishes nicely between the three trees with
the correct tree tightly distributed around zero.  It is correct 
almost all of the time.
Lake's linear invariant (middle)
also shows power to distinguish between all three trees, but
distributions overlap much more---it is only correct about half of the 
time.  The final polynomial seems to be biased towards
selecting the wrong tree, even though it does not lie in $I(T)$ for either of the other two trees.

\begin{figure}
	\centering
	\includegraphics[width=4in]{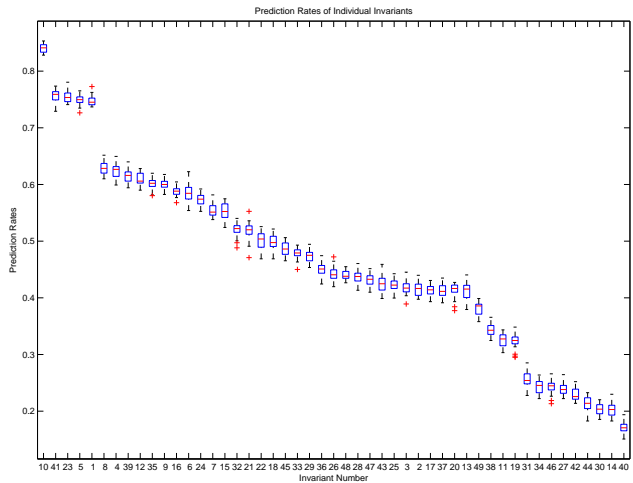}
	\caption{Prediction rate for the 49 Jukes-Cantor invariants on simulated data of length 100.  
	The four-point invariant is by far the best, although four other invariants are quite good.}
	\label{fig:ranks}
\end{figure}

Figure~\ref{fig:ranks} shows the performance of all the generators for this
ideal on simulated data.  The four-point invariant is the best, but the
performance drops sharply with the other generators.  Notably, the four-point
invariant and several of the other powerful ones are unchanged (aside from
sign) under the symmetries of the tree.  While  any invariant can be made
symmetric by averaging, this behavior leads us to believe that invariants with
a simple, symmetric form may be the best choice.

For more complex models, it becomes even more necessary to pick a good set of
invariants since there are prohibitively many generators of the ideal.  The paper
\cite{Casanellas2007} describes an algebraic method for picking a subset of
invariants for the Kimura 3-parameter model, which has 11612 generators for the
quartet tree (after augmenting by symmetry).  Their method constructs a set of
invariants which is a local complete intersection, and shows that this 
defines the variety on the biological relevant region.  This reduces the
list to 48 invariants which overall behave better than all 11612 invariants.
However, of these 48, only 4 rank among the top 52  invariants in
prediction rate (using simulations similar to those that  produced
Figure~\ref{fig:ranks}) and the remaining 44 invariants are mostly quite poor
(42\% average accuracy).  This
result, while of considerable theoretical interest, doesn't seem to give an
optimal set of invariants.


\section{Comparing trees} \label{sec:compare}
Once we have chosen a set $\bfT$ of invariants for each tree $T$, we want to
pick the tree such that $\bfT(\hat{p})$ is smallest (in some sense).
The examples in Section~\ref{sec:choose} show why this is a non-trivial problem---different 
invariants have different power and different variance and thus
should be weighted differently in choosing a norm on $\bfT$.
In this section, we briefly describe an approach to normalizing the invariants
to enable us to choose a tree.  It is based on machine learning and was
developed in \cite{Eriksson2007}.  It leads to large improvements over previous
uses of invariants; however, it is computationally expensive  and dependant on
the training data.
It can be thought of as finding the best single invariant which is a quadratic
form in the starting set $\bfT$ of invariants.

There are also standard asymptotic statistical tools such as the delta method
for normalizing invariants to have a common mean and variance.  They have the
disadvantage of depending on a linear approximation and asymptotic behavior,
which might not be accurate for small datasets.  Fortunately, the varieties for
many phylogenetic models are smooth in the biologically significant region
\cite{Casanellas2007}, so linear approximations may work well.

This problem is somewhat easier when we are choosing between different
trees with the same (unlabelled) topology, for example, the three quartet trees.  In this
case the different ideals $I(T)$ are the same under a permutation of the unknowns, 
and thus we are comparing
the same sets of polynomials (as long as the chosen set $\bfT$ is closed under
the symmetries of $T$).  For this reason, we will concentrate on the case of
quartet trees and 
write $T_1 = (01 : 23)$, $T_2 = (02 : 13)$, and $T_3 = (03 : 12)$.

Let $\hpt$ be an empirical probability distribution generated from a
phylogenetic model on tree $T_1$ with parameters $\theta$. 
Choose $m$ invariants $\bof_i$ ($i=1,2,3$) that are closed under the
symmetries of $T_1$.  We want a norm $\|\ \|_\ast$ such
that
\begin{equation}
	\label{eq:cond}
	\|\bof_1(\hpt))\|_\ast < \min\left( \|\bof_2(\hpt)\|_\ast, \|\bof_3(\hpt)\|_\ast \right)
\end{equation}
is typically true, i.e., the true tree should have its associated invariants
closer to zero than others on the relevant range of parameter space. 

In order to scale and weigh the
individual invariants, the algorithm seeks to find an optimal 
$\|\ \|_\ast$ within the class of Mahalanobis norms. Recall that given a
positive (semi)definite matrix $A$, the Mahalanobis (semi)norm $\|\cdot \|_A$
is defined by 
\[
\| x \|_A = \sqrt{x^t A x}.
\]
Since $A$ is positive semidefinite, it can be written as $A = U D
U^t$ where $U$ is  orthogonal and $D$ is diagonal with non-negative entries.
Thus the positive semidefinite square root $B = U \sqrt{D} U^t$ is unique.  Now since $\|x\|_A^2 =
x^tAx = (Bx)^t(Bx)=\|Bx\|^2$,  learning such a metric is the same as finding a
transformation of the space of invariants that replaces each point $x$ with
$Bx$ under the Euclidean norm, i.e., a rotation and shrinking/stretching of the
original $x$.  

Now suppose that $\Theta$ is a finite set of parameters from which 
training data $\bof_1(\hpt), \bof_2(\hpt), \bof_3(\hpt)$ is generated
for $\theta \in \Theta$.
As we saw above, each of the eight possible ways of writing
each tree induces a permutation of the coordinates $p_{ijkl}$ and thus induces
a signed permutation of the coordinates of each $\bof_i(\hpt)$.  Write these
permutations in matrix form as $\pi_1, \dots,
\pi_8$.
Then the positive semidefinite matrix $A$ must satisfy the symmetry 
constraints $\pi_i A = A \pi_i$ which are hyperplanes intersecting the
semidefinite cone.  This symmetry condition is crucial in  reducing the
computational cost.
Given training data, the following optimization problem finds a good metric
on the space of invariants.
\begin{equation}
	\label{eq:metlearn}
	\begin{array}{ll}
		\text{Minimize: } &\sum_{\theta \in \Theta} \xi(\theta) + \lambda \tr A\\
		\text{Subject to: } \quad
		&\| \bof_1 (\hpt) \|_A^2   + \gamma \leq  \| \bof_i (\hpt)\|_A^2 + \xi(\theta)\quad (\text{for } i=2,3), \\
		& \pi_i A = A \pi_i \quad(\text{for } 1 \leq i \leq 8),\\
		&\xi(\theta)  \geq 0, \quad\text{and}\\
		&A \succeq 0,
	\end{array}
\end{equation} 
where $A\succeq 0$ denotes that $A$ is a positive semidefinite matrix, so 
this is a semidefinite programming problem.  
There are several parameters involved in this algorithm: $\xi(\theta)$ for
$\theta\in \Theta$ are slack-variables measuring the violation of
(\ref{eq:cond}), $\gamma$ is a margin parameter that lets us strengthen
condition (\ref{eq:cond}), and $\lambda$ is a regularization parameter to keep
the trace $\tr A$ small while keeping $A$ as low rank as possible. It tries to
find a positive semidefinite $A$ at a trade-off between the small violation of
(\ref{eq:cond}) and small trace $A$.  

The metric learning problem (\ref{eq:metlearn}) was inspired by some early
results on metric learning algorithms \cite{Xing2003,Shalev-Shwartz2004}, which
aim to find a Mahalanobis (semi)norm such that the mutual distances between
similar examples are minimized while the distances across dissimilar examples
or classes are kept large.  If it becomes too computationally expensive, we can
restrict $A$ to be diagonal, which reduces the problem to a linear program.
See \cite{Eriksson2007} for details and simulation results.  The learned
metrics significantly improve on any of the individual invariants as well as on
unweighted norms.  The semidefinite programming algorithm is computationally
feasible for approximately 100 invariants, and the choice of powerful
invariants is important.

\section{Efficient computation} \label{sec:compute}

At first glance, the problem of using invariants seems intractable for large
trees for the simple reason that the number of unknowns  grows
exponentially with the number of leaves.  However, the problem is not as bad as
it may seem.  Phylogenetic analyses typically use DNA sequences at most
thousands of bases long, which means that the empirical distribution $\hat{p}
\in \rr^{4^n}$ will be extremely sparse even with a relatively small number of
taxa.

Also the data can be sparse, this will not help unless we can write down the invariants
in sparse form.  
If the polynomials can be written down in an effective way, they can be
evaluated quickly.  The determinantal form of the invariants in
Theorem~\ref{thm:flat}  provide such a form; see \cite{Eriksson2005} for an
algorithm to construct phylogenetic trees in polynomial time using these
invariants and numerical linear algebra.  Also see \cite{Allman2003}
for invariants that are (in some sense) determinantal.
It seems that determinantal
conditions could be particularly useful, so we suggest Problem~\ref{prob:det}
to computational commutative algebraists (see also \cite{Eriksson2005a}).

Unfortunately, for group-based models the polynomials are only sparse when written in
Fourier coordinates, and the Fourier transform takes a sparse distribution $p$
and produces a completely dense vector $q$.  Many of the invariants are
determinantal in Fourier coordinates, but since the matrices are dense, they
are difficult to write down.  Can these polynomials be evaluated efficiently?

\section{Positivity} \label{sec:pos}

Recall that the four point condition 
(Proposition~\ref{prop:4pt} and Figure~\ref{fig:4pt})
says that
for a dissimilarity map $d$ arising from the quartet tree $(01 : 23)$,
\begin{equation}
	d_{01} + d_{23} \leq d_{02} + d_{13} = d_{03} + d_{12}.
	\label{eq:4p}
\end{equation}
This is true since the right two sums traverse the inner edge of the tree twice 
(Figure~\ref{fig:4pt}).
We saw in Example~\ref{ex:4pt} that the equality in (\ref{eq:4p}) translates to
a quadratic invariant.  However, notice that if the interior branch of the
tree has  negative length, the equality is still satisfied, but the inequality changes
so that $d_{01} + d_{23}$ is now larger than the other two sums.

The widely used neighbor-joining algorithm \cite{Saitou1987}, when restricted
to four taxa, reduces to finding the smallest of the three sums in the
four-point condition.  That is, neighbor-joining on a quartet tree
involves estimating the
distances as in Section~\ref{sec:back} and then returning the tree $(ij :
kl)$ that minimizes $d_{ij} + d_{kl}$.
If instead we used the quadratic invariant arising from the equality in
the four point condition, we would 
have an invariant based method that simply returns the tree $(ij : kl)$ that 
minimizes $|d_{ik} + d_{jl} - d_{il} - d_{jk}|$.  We saw in
Section~\ref{sec:choose} that this invariant performs quite well compared to
the other generators of the Jukes-Cantor model.  However, it compares poorly
to the neighbor-joining criterion in the following way.

\begin{figure}
	\centering
	\includegraphics[width=.85\textwidth]{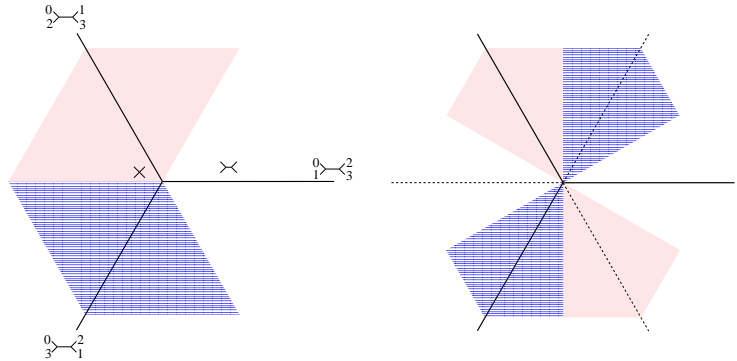}
	\caption{The selection criteria for neighbor-joining (left) and the
	four-point invariant (right) projected to two dimensions.  The
	colored/shaded regions show which dissimilarity maps are matched to which
	trees by the two algorithms.  The white/unshaded area corresponds to tree
	$(01 : 23)$, the red/solid area to tree $(02 : 13)$ and the blue/striped
	area to $(03 : 12)$.} 
	\label{fig:dist}
\end{figure}

Figure~\ref{fig:dist} shows the difference between these two selection criteria
on a projection of the six dimensional space of dissimilarity maps
$\rr^{\binom{4}{2}}$ to two dimensions.  
The three  black lines are the projections of distances arising from the
three different trees.   Moving out from the center along these lines
corresponds to increasing the length of the inner edge in the tree.

Geometrically, neighbor-joining can be thought of as finding the closest tree  metric
(a point on a black half-ray) to a dissimilarity map.  
The four-point invariant can't distinguish negative inner
branch length (the dotted black line) and thus is much less robust than
neighbor-joining.  Notice that even when it picks the wrong tree, it can pick
the \emph{wrong} wrong tree---that is, the one least supported by the data.
It is less robust than neighbor-joining in the ``Felsenstein zone''
\cite{Huelsenbeck1993}, which corresponds to the region close to the center, where
the inner edge is very short.

\begin{figure}
	\centering
	\begin{tabular}{cc}
		\includegraphics[width=.43\textwidth]{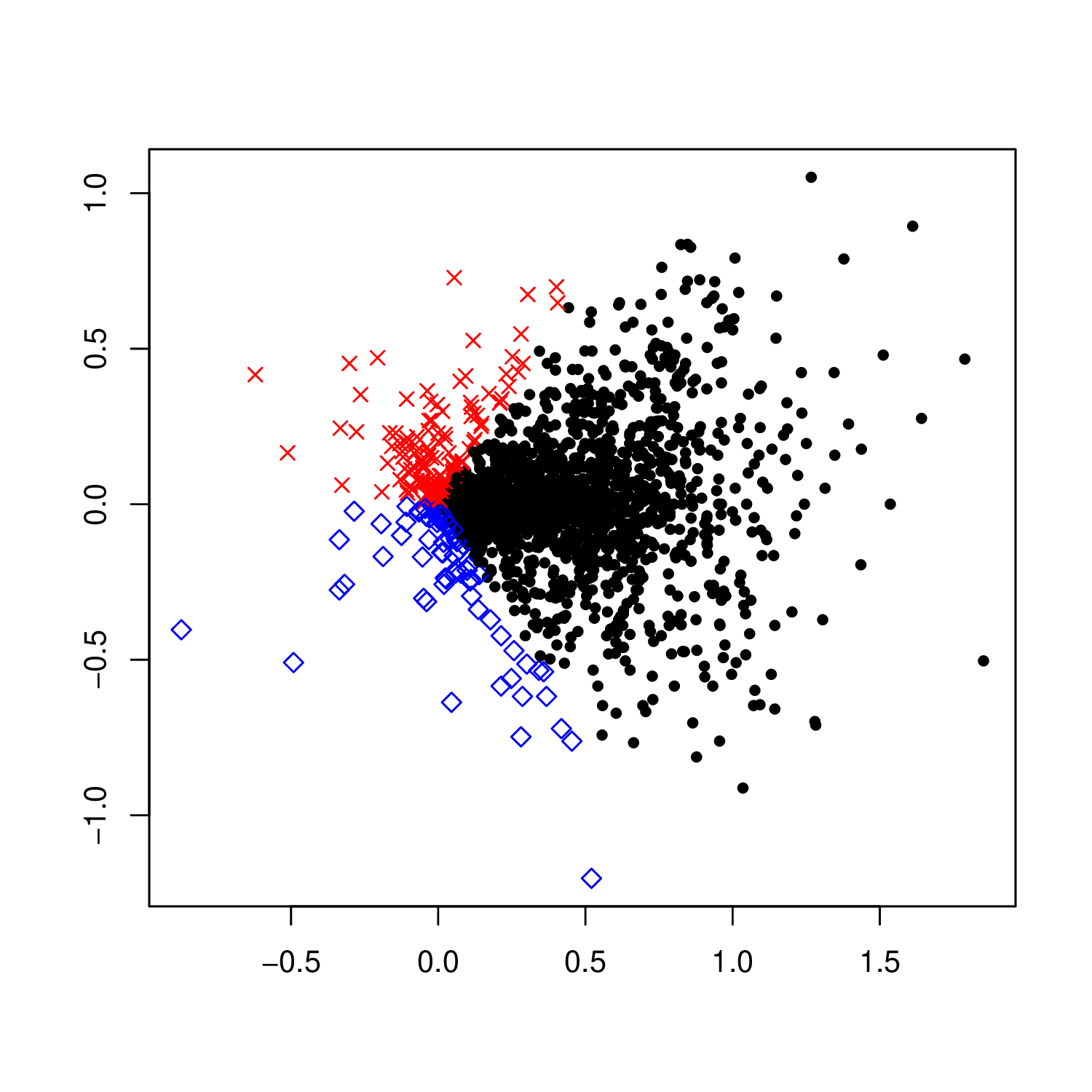} & 
		\includegraphics[width=.43\textwidth]{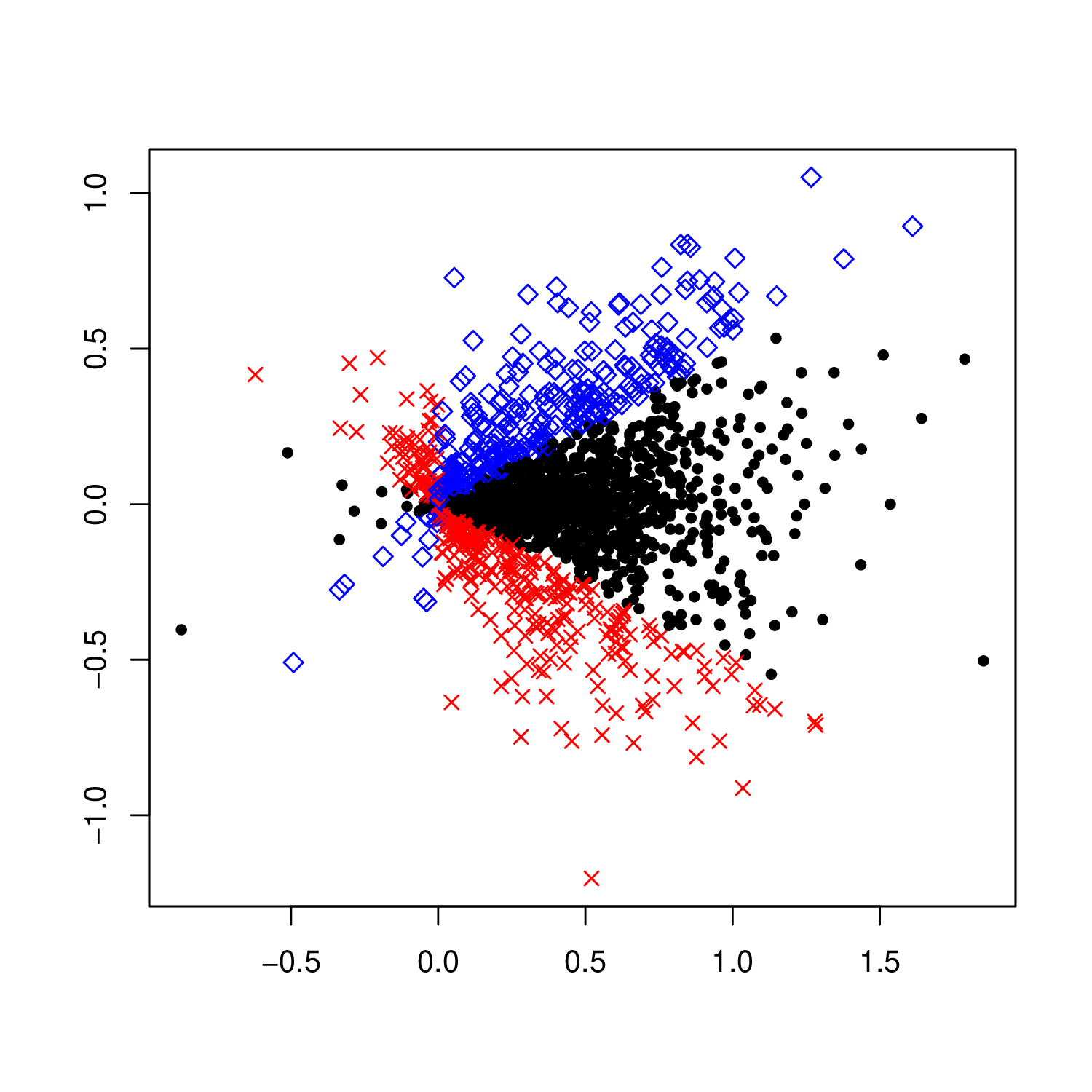} \\
		Neighbor joining & Four-point invariant\\
	\end{tabular}
	\caption{Illustration of Figure~\ref{fig:dist} on simulated data.
	Simulated alignments from tree $(01 : 23)$ of length 100 were created for
	randomly chosen branch lengths between $0.01$ and $0.75$.  Distances were
	estimated using the Jukes-Cantor model and projected onto two dimensions in
	the same way as in Figure~\ref{fig:dist}.  Trees were built from the
	distances using both neighbor-joining and the four-point invariant. Black
	circles correspond to distances assigned tree $(01 : 23)$, red x's to tree
	$(02 : 13)$, and blue diamonds to tree $(03 : 12)$.}
	\label{fig:distR}
\end{figure}

Simulations (see Figure~\ref{fig:distR}) show that
building trees by evaluating this quadratic invariant 
does not perform nearly as well as neighbor-joining.  This is
because many simulations with a short interior branch 
tend to return metrics that seem to come from trees with negative inner branch lengths.

This seems to be a large blow to the method of invariants: even the most
powerful invariant on our list in Section~\ref{sec:choose} doesn't behave as
well as this simple condition.  However, it can be easily seen that testing the
inequality is equivalent to testing the signs of the invariant instead of the
absolute value, which leads us to ask if invariants can provide a way to
discover conditions similar to that used in neighbor-joining (see
Problem~\ref{question:4pt}).  The original paper of Cavender-Felsenstein
\cite{Cavender1987} also suggested using inequalities, although no one seems to have
followed up on this idea.

\section{Open problems} \label{sec:prob}

\begin{question}\label{prob:length}
	Can algebraic ideas be used to estimate branch lengths and other parameters
	in phylogenetic trees?  See \cite{Steel2000, Allman2008} for algebraic
	techniques for estimating parameters in invariable-site phylogenetic
	models.
\end{question}

\begin{question} \label{prob:hetero}
	Investigate the behavior of individual invariants on data from trees with
	heterogeneous rates.  Are the best invariants the same ones that are
	powerful for homogeneous rates?
\end{question}


\begin{question}
	Can asymptotic statistical methods be practically used to normalize
	invariants?  Do they give any information about the power of individual
	invariants?
\end{question}

\begin{question} 
	Do the metrics constructed by the machine learning algorithm
	in Section~\ref{sec:compare} shed any light on the criteria for
	invariants to be powerful?
\end{question}

\begin{question}\label{prob:det}
	Define the ``determinantal closure'' of an ideal $I$ and develop algorithms
	to calculate it. See also \cite{Eriksson2005a}.
\end{question}

\begin{question} For group-based models, does Fourier analysis provide a method
	to efficiently evaluate polynomials in the Fourier coordinates without
	destroying the sparsity of the problem?  Note that many of the invariants
	are determinental in Fourier coordinates.  
\end{question}

\begin{question}\label{question:4pt}
	Are there other phylogenetic invariants (say for quartet
	trees under the Jukes-Cantor model) ``similar'' to the four-point invariant?
	We suggest the following conditions:
	\begin{enumerate}
		\item Be fixed (up to sign) under the $\zz_2 \times \zz_2 \times \zz_2$
			symmetries of the quartet tree.
		\item Have the following sign condition: $\pm f(p) > 0$ for all $p$ from 
			$T_2$ and $T_3$ (with perhaps a different choice of sign for $T_2$ and $T_3$).
			See for example, the symmetries of the left subfigure in
			Figure~\ref{fig:hist}.  
	\end{enumerate}
	 Beware that  results such as \cite{Bryant2005} on the uniqueness of the
	 neighbor-joining criterion place some constraints on whether we can hope
	 to find invariants mimicking this behavior.
\end{question}
\section*{Acknowledgments}
We thank E.~Allman, M.~Casanellas, M.~Drton, L.~Pachter, J.~Rhodes, and
F.~Sottile for enlightening discussions about these topics at the IMA.  We are
very appreciative of the hospitality of the IMA during our visit and of very
helpful comments from two referees.

\bibliographystyle{siam}

\end{document}